\documentclass[longauth]{aa}

\usepackage[varg]{txfonts}
\usepackage{graphicx}
\usepackage{lscape}
\usepackage{amsmath}
\usepackage{amssymb}
\usepackage{color}
\usepackage{natbib}
\bibpunct{(}{)}{;}{a}{}{,} 
\usepackage{url}
\usepackage{multirow}
\usepackage{threeparttable}  
\usepackage{booktabs} 
\usepackage{soul}
\usepackage{float}
\usepackage[colorlinks=true, linkcolor=blue, citecolor=blue, urlcolor=blue]{hyperref}
\usepackage{tablefootnote}
\usepackage{subcaption}
\usepackage{academicons}
\usepackage{xcolor}
\usepackage{orcidlink}
\usepackage{multicol}
\usepackage{booktabs}
\usepackage{longtable}
\usepackage{caption}
\usepackage{lastpage}
\hypersetup{colorlinks=true,allcolors=blue,citecolor=blue}
\usepackage[font=small,labelfont=bf]{caption}
\usepackage{siunitx}

\newcommand{\arcs}{$^{\prime\prime}$}
\authorrunning{Srivastava et al.}

\begin{document}

   \title{Night-sky emission correction techniques for high-resolution spectroscopy: Demonstration with NIRPS}

\author{
Avidaan Srivastava\inst{1,2,*}\orcidlink{0009-0009-7136-1528},
Fran\c{c}ois Bouchy\inst{2}\orcidlink{0000-0002-7613-393X},
Xavier Dumusque\inst{2}\orcidlink{0000-0002-9332-2011},
\'Etienne Artigau\inst{1,3}\orcidlink{0000-0003-3506-5667},
Ren\'e Doyon\inst{1,3}\orcidlink{0000-0001-5485-4675},
Neil J. Cook\inst{1}\orcidlink{0000-0003-4166-4121},
Danuta Sosnowska\inst{2},
Flavie B\'elanger\inst{1}\orcidlink{0009-0000-3012-8337},
David Lafreni\`ere\inst{1}\orcidlink{0000-0002-6780-4252},
Fr\'ed\'erique Baron\inst{1,3}\orcidlink{0000-0002-5074-1128},
Lison Malo\inst{1,3},
Susana C. C. Barros\inst{4,5}\orcidlink{0000-0003-2434-3625},
Bj\"orn Benneke\inst{6,1}\orcidlink{0000-0001-5578-1498},
Xavier Bonfils\inst{7}\orcidlink{0000-0001-9003-8894},
Marta Bryan\inst{8},
Ryan Cloutier\inst{9}\orcidlink{0000-0001-5383-9393},
Nicolas B. Cowan\inst{10,11}\orcidlink{0000-0001-6129-5699},
Elisa Delgado-Mena\inst{12,4}\orcidlink{0000-0003-4434-2195},
Xavier Delfosse\inst{7}\orcidlink{0000-0001-5099-7978},
David Ehrenreich\inst{2,13},
Pedro Figueira\inst{2,4},
Jonay I. Gonz\'alez Hern\'andez\inst{14,15}\orcidlink{0000-0002-0264-7356},
Izan de Castro Le\~ao\inst{16}\orcidlink{0000-0001-5845-947X},
Christophe Lovis\inst{2}\orcidlink{0000-0001-7120-5837},
Bruno L. Canto Martins\inst{16}\orcidlink{0000-0001-5578-7400},
Lucile Mignon\inst{2,7},
Christoph Mordasini\inst{17}\orcidlink{0000-0002-1013-2811},
Francesco Pepe\inst{2}\orcidlink{0000-0002-9815-773X},
Rafael Rebolo\inst{14,15,18}\orcidlink{0000-0003-3767-7085},
Jason Rowe\inst{19},
Nuno C. Santos\inst{4,5}\orcidlink{0000-0003-4422-2919},
Damien S\'egransan\inst{2},
Alejandro Su\'arez Mascare\~no\inst{14,15}\orcidlink{0000-0002-3814-5323},
Jose Renan De Medeiros\inst{16}\orcidlink{0000-0001-8218-1586},
St\'ephane Udry\inst{2}\orcidlink{0000-0001-7576-6236},
Diana Valencia\inst{8}\orcidlink{0000-0003-3993-4030},
Gregg Wade\inst{20,21},
Romain Allart\inst{1}\orcidlink{0000-0002-1199-9759},
Vincent Bourrier\inst{2}\orcidlink{0000-0002-9148-034X},
Pedro Branco\inst{5,4}\orcidlink{0009-0007-5130-5188},
Charles Cadieux\inst{1}\orcidlink{0000-0001-9291-5555},
Gaspare Lo Curto\inst{22}\orcidlink{0000-0002-1158-9354},
Yolanda G. C. Frensch\inst{2,22,}\orcidlink{0000-0003-4009-0330},
Jonathan Gagn\'e\inst{23,1},
Roseane de Lima Gomes\inst{1,16}\orcidlink{0000-0002-2023-7641},
Nicole Gromek\inst{9}\orcidlink{0009-0000-1424-7694},
Vigneshwaran Krishnamurthy\inst{10}\orcidlink{0000-0003-2310-9415},
Pierrot Lamontagne\inst{1},
Yuri S. Messias\inst{1,16}\orcidlink{0000-0002-2425-801X},
Khaled Al Moulla\inst{4,2}\orcidlink{0000-0002-3212-5778},
Dany Mounzer\inst{2}\orcidlink{0000-0002-8070-2058},
Nicola Nari\inst{24,14,15},
L\'ena Parc\inst{2}\orcidlink{0000-0002-7382-1913},
Atanas K. Stefanov\inst{14,15}\orcidlink{0000-0002-6059-1178},
Thomas Vandal\inst{1}\orcidlink{0000-0002-5922-8267},
Drew Weisserman\inst{9}\orcidlink{0000-0002-7992-469X},
Joost P. Wardenier\inst{1}\orcidlink{0000-0003-3191-2486}
}

\institute{
\inst{1}Institut Trottier de recherche sur les exoplan\`etes, D\'epartement de Physique, Universit\'e de Montr\'eal, Montr\'eal, Qu\'ebec, Canada\\
\inst{2}Observatoire de Gen\`eve, D\'epartement d’Astronomie, Universit\'e de Gen\`eve, Chemin Pegasi 51, 1290 Versoix, Switzerland\\
\inst{3}Observatoire du Mont-M\'egantic, Qu\'ebec, Canada\\
\inst{4}Instituto de Astrof\'isica e Ci\^encias do Espa\c{c}o, Universidade do Porto, CAUP, Rua das Estrelas, 4150-762 Porto, Portugal\\
\inst{5}Departamento de F\'isica e Astronomia, Faculdade de Ci\^encias, Universidade do Porto, Rua do Campo Alegre, 4169-007 Porto, Portugal\\
\inst{6}Department of Earth, Planetary, and Space Sciences, University of California, Los Angeles, CA 90095, USA\\
\inst{7}Univ. Grenoble Alpes, CNRS, IPAG, F-38000 Grenoble, France\\
\inst{8}Department of Physics, University of Toronto, Toronto, ON M5S 3H4, Canada\\
\inst{9}Department of Physics \& Astronomy, McMaster University, 1280 Main St W, Hamilton, ON, L8S 4L8, Canada\\
\inst{10}Department of Physics, McGill University, 3600 rue University, Montr\'eal, QC, H3A 2T8, Canada\\
\inst{11}Department of Earth \& Planetary Sciences, McGill University, 3450 rue University, Montr\'eal, QC, H3A 0E8, Canada\\
\inst{12}Centro de Astrobiolog\'ia (CAB), CSIC-INTA, Camino Bajo del Castillo s/n, 28692, Villanueva de la Ca\~nada (Madrid), Spain\\
\inst{13}Centre Vie dans l’Univers, Facult\'e des sciences de l’Universit\'e de Gen\`eve, Quai Ernest-Ansermet 30, 1205 Geneva, Switzerland\\
\inst{14}Instituto de Astrof\'isica de Canarias (IAC), Calle V\'ia L\'actea s/n, 38205 La Laguna, Tenerife, Spain\\
\inst{15}Departamento de Astrof\'isica, Universidad de La Laguna (ULL), 38206 La Laguna, Tenerife, Spain\\
\inst{16}Departamento de F\'isica Te\'orica e Experimental, Universidade Federal do Rio Grande do Norte, Campus Universit\'ario, Natal, RN, 59072-970, Brazil\\
\inst{17}Space Research and Planetary Sciences, Physics Institute, University of Bern, Gesellschaftsstrasse 6, 3012 Bern, Switzerland\\
\inst{18}Consejo Superior de Investigaciones Cient\'ificas (CSIC), E-28006 Madrid, Spain\\
\inst{19}Bishop's Univeristy, Dept of Physics and Astronomy, Johnson-104E, 2600 College Street, Sherbrooke, QC, Canada, J1M 1Z7, Canada\\
\inst{20}Department of Physics, Engineering Physics, and Astronomy, Queen’s University, 99 University Avenue, Kingston, ON K7L 3N6, Canada\\
\inst{21}Department of Physics and Space Science, Royal Military College of Canada, 13 General Crerar Cres., Kingston, ON K7P 2M3, Canada\\
\inst{22}European Southern Observatory (ESO), Av. Alonso de Cordova 3107,  Casilla 19001, Santiago de Chile, Chile\\
\inst{23}Plan\'etarium de Montr\'eal, Espace pour la Vie, 4801 av. Pierre-de Coubertin, Montr\'eal, Qu\'ebec, Canada\\
\inst{24}Light Bridges S.L., Observatorio del Teide, Carretera del Observatorio, s/n Guimar, 38500, Tenerife, Canarias, Spain\\
\inst{*}\email{avidaan.srivastava@umontreal.ca}
}

   \date{XXX 2025}

\abstract
    {Ground-based spectrographs operating in the near-infrared (NIR) regime are hampered by various absorption and emission features of Earth’s atmosphere. While considerable attention has been paid to mitigating telluric absorption, correcting telluric emission features remains non-trivial and can significantly affect the observation of faint targets.}
    {We aim to develop and implement automated algorithms for sky background emission correction in the context of high-resolution spectroscopy. These empirical-based algorithms have been officially integrated into both NIRPS data reduction pipelines: NIRPS DRS and APERO DRS. The NIRPS DRS algorithm is additionally functional at the telescope site, enabling quick, real-time telluric-corrected data processing. Designed for flexibility, these techniques can be incorporated into the reduction workflow of any high-resolution spectrograph to improve the radial velocity (RV) performance.}
    {In our approach, a reference sky spectrum is first created by deep-stacking NIRPS sky calibration frames on a common wavelength grid and calculating the weighted median flux per pixel, separately for both the high-accuracy and high-efficiency instrument modes. This process is repeated for all spectral orders and for both the object  and sky-calibration fibres: fibre A and fibre B, respectively. In this reference sky spectrum, the sky emission lines can be identified and used to construct a static library. During the data reduction process, the emission features in the library are individually scaled in terms of flux using two distinct techniques, each specific to the two DRS pipelines. Finally, they are locally subtracted from the science observations to minimise their noise contribution to the final spectrum}
    {We find that the correction algorithms significantly improve the RV measurements obtained using both the cross-correlation function and line-by-line techniques, enabling NIRPS to achieve submetre-per-second precision in the NIR. The techniques have been successfully validated and demonstrated on both bright (Proxima Centauri) and faint (TOI-406, TOI-3494, TOI-4552) targets, as well as in both high-efficiency and high-accuracy instrument modes.}
    {We present the first attempts at empirically subtracting emission line features from high-resolution spectra. These algorithms can be applied independently of the instrumentation and implemented into the data reduction workflow of any high-resolution instrument with minimal modifications required.}

\keywords{sky emission --
            OH molecules --
            night sky airglow --
            near-infrared spectroscopy
           }

\maketitle

\section{Introduction}

The majority of stars in the galaxy are not G-type stars such as the Sun. Instead, M dwarfs tend to dominate \citep[e.g.][]{Chabrier2001, bochanski2010}, which are cooler ($T_{\rm eff}<3800$\,K), smaller, and less massive (by $\sim$10-50\%) relative to the Sun. These stars host a large number of rocky planets \citep[e.g.][]{Bonfils2013, Dressing&Charbonneau2015} and their spectral energy distribution (SED) peaks in the near infrared (NIR) rather than the visible (VIS) wavelength. To capitalise on these exoplanet detections, new NIR instruments have been developed, such as the Near Infra-Red Planet Searcher (NIRPS, \citealt{nirps, Bouchy2025}). NIRPS is a high-resolution, fibre-fed, adaptive optics assisted, echelle spectrograph that operates in the NIR wavelength domain of 0.98\,$\mu$m to 1.9\,$\mu$m ($YJH$). It is mounted on the ESO 3.6\,m telescope in La Silla Observatory and is designed to work alongside the High Accuracy Radial velocity Planet Searcher (HARPS;~\citealt{harps}) spectrograph to provide full VIS and NIR wavelength coverage. NIRPS operates in a thermally controlled environment inside a vacuum vessel, making it highly stable and offers two instrument modes with differing spectral resolutions: high accuracy (HA, $\lambda/\Delta \lambda \sim 85\,000$) and high efficiency (HE, $\lambda/\Delta \lambda \sim 75\,000$). Observations in both modes make use of two fibres: one centred on the target star and the other to serve as a simultaneous calibration either on the sky background (possible due to the high stability of the instrument) or the Fabry-Pérot etalon.

Processed data from NIRPS can be obtained through two independent data reduction softwares (DRS) that are both capable of processing and reducing the raw spectrum into science-ready 2D and 1D spectra. Here, the 2D spectrum refers to a 2D array (or fits image) containing the wavelength-calibrated flux from all individual echelle spectral orders dispersed across the detector. On the other hand, the 1D spectrum is a so-called wavelength-stitched deblazed spectrum that combines all the detected spectral orders of the 2D spectrum, thereby covering the entire wavelength domain of the instrument. Although they are functionally distinct, the presence of two DRS pipelines allows us to simultaneously optimise their performance, evaluate their respective strengths and weaknesses, and efficiently compare the reduced products to correct any discrepancies or errors. The ESO-supported DRS that functions at the telescope is a modified version of the publicly accessible ESPRESSO pipeline (hereby, NIRPS DRS\footnote{NIRPS User Manual, NIRPS-2000-GEN-UM-148, Issue 2.3, https://www.eso.org/sci/facilities/lasilla/instruments/nirps/doc.html}), capable of functioning in the NIR regime. An alternative DRS pipeline is A PipelinE to Reduce Observations (APERO), an initial re-coding of the HARPS pipeline, developed in the APERO framework \citep{apero} as the official DRS for the SPectropolarimètre InfraRouge (SPIRou, \citealt{spirou2}) spectrograph, but has now been adapted to work on NIRPS data. The NIRPS DRS extracts 71 spectral orders, while the APERO DRS extracts 75, with the 4 extra orders being very weak in flux as they lie between atmospheric windows and, hence, they do not end up contributing to the radial velocity (RV) precision. For computational efficiency, RV measurements are obtained at the telescope site using the traditional cross-correlation function (CCF, \citealt{baranne96, pepe2002}) method, while extreme precision RV (EPRV) tools such as line-by-line (LBL, \citealt{lbl}) analysis are used in post-processing to reach performances at the 1\,m/s RV limit or better.

This work is focussed on the development of automatic, empirical-based algorithms to correct for the non-thermal sky background emission due to the airglow of Earth's atmosphere prevalent in the NIR. The principles of the algorithms can be applied to any high-resolution spectrograph (e.g. ANDES at ELT;~\citealt{marconi2022}) and the removal of these features, along with telluric absorption lines \citep{artigau2014, allart2022}, is imperative for $\le$1~m/s
precision RV measurements needed to detect Earth-like planets. The paper is structured as follows: Section \ref{sec2} discusses the properties of telluric emission and current methods in literature used to mitigate their effects. Section \ref{sec3} describes the specific approaches used in NIRPS for the telluric emission correction in both the NIRPS DRS and APERO DRS. Section \ref{sec4} focusses on the various tests (spectrum level and impact on RV) done to validate our methods. Section \ref{sec5} discusses the implementation of the correction algorithm in the context of the DRS workflow, along with methods to remove any telluric residuals post-correction that might still affect RV measurements. Finally, the conclusions of this work are presented in Section \ref{sec6}.

\section{Telluric emission in NIR wavelengths} \label{sec2}

\subsection{Properties of OH and O$_2$ lines as NIR sky emission}

Studies of the night sky airglow (or emission spectrum) go as far back as the 1950s \citep{meinel1950, bates1950, anlauf1968}. By obtaining, for the time, high-resolution IR spectra of the sky, a consensus was reached that it was primarily the rotational-vibrational modes of the OH radical that were responsible for the airglow. In the upper layers of the atmosphere (in the ozone layer), water vapour is bombarded by ultraviolet photons from the Sun and dissociates into hydrogen and hydroxyl ions. These lone hydrogen atoms then collide with ozone molecules. After several intermediate steps, they produce hydroxyl ions following

\begin{equation} \label{oheqn} 
    H\, +\, O_{3}\, \xrightarrow{} OH_{\nu \le 9}\, +\, O_2, 
\end{equation}

where $\nu$ refers to the vibrational modes. These vibrational modes get excited during the creation of the OH, thereby resulting in IR emission as they get de-excited. In addition, molecular oxygen, produced in the same chemical reaction also emits in the small NIR band of 1.25\,$\mu$m to 1.3\,$\mu$m \citep{kmsmith1999}. 

These emission lines are numerous and populate the entire wavelength regime of NIRPS as in Fig. \ref{s1d_emission}. The flux here, and in all following figures is recorded in photo-electrons per second. A theoretically derived library of the night sky airglow was presented in \citet{rousselot2000}. However, variations in intensities of lines originating from different vibrational modes and the limited spectral resolution of the instrument mean that not all lines can be detected or resolved. To address this, we compiled a custom list of lines to correct, based on those identified in NIRPS night-sky exposures. This custom line list, in the form of the reference sky (section \ref{ref_sky}) spectrum, is overplotted in Fig. \ref{s1d_emission} to demonstrate that the majority of lines are detected and corrected using our algorithms. 

\begin{figure}
\includegraphics[width=\columnwidth]{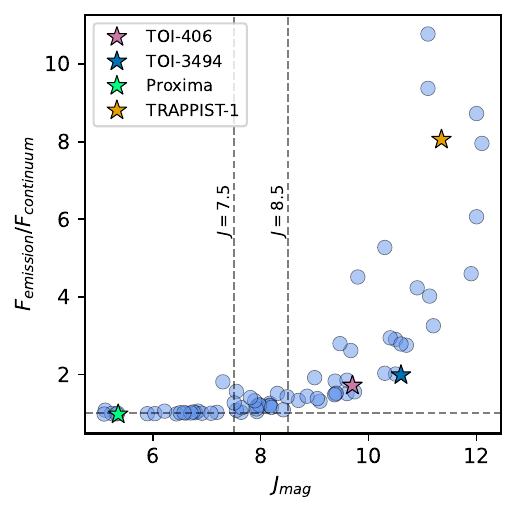}
\centering
\caption[]{Observed stellar flux at the location of a known strong emission line doublet at 1.603055 $\mu$m ($F_{emission}$) is compared to the surrounding continuum ($F_{continuum}$) flux as a function of magnitude in the \textit{J} band. At $J\sim7.5$, the ratio begins to slightly deviate from unity, indicating that the emission features are no longer blended into the continuum beyond this magnitude. Notable stars are highlighted for reference.}
\label{emission_jmag}
\end{figure}

For the closest bright stars, this airglow is relatively negligible as the observed flux is dominated by the stellar continuum. However, for faint targets ($J$ mag > 8.5), the background airglow begin to contaminate the spectrum to such an extent and if it is uncorrected for, it can skew RV measurements. Fig.~\ref{emission_jmag} demonstrates this effect by comparing the flux at the locality of a strong OH emission line at 1.603055\,~$\mu$m with the immediate surrounding continuum across a variety of targets observed by NIRPS. The ratio is unity for the bright targets and deviates from unity exponentially as the magnitude of the target increases starting at approximately $J\sim$\,8.5. Some notable bright (Proxima Centauri) and faint (TOI-406, TOI-3494, TRAPPIST-1) are highlighted for reference.

\begin{figure}
\includegraphics[width=\columnwidth]{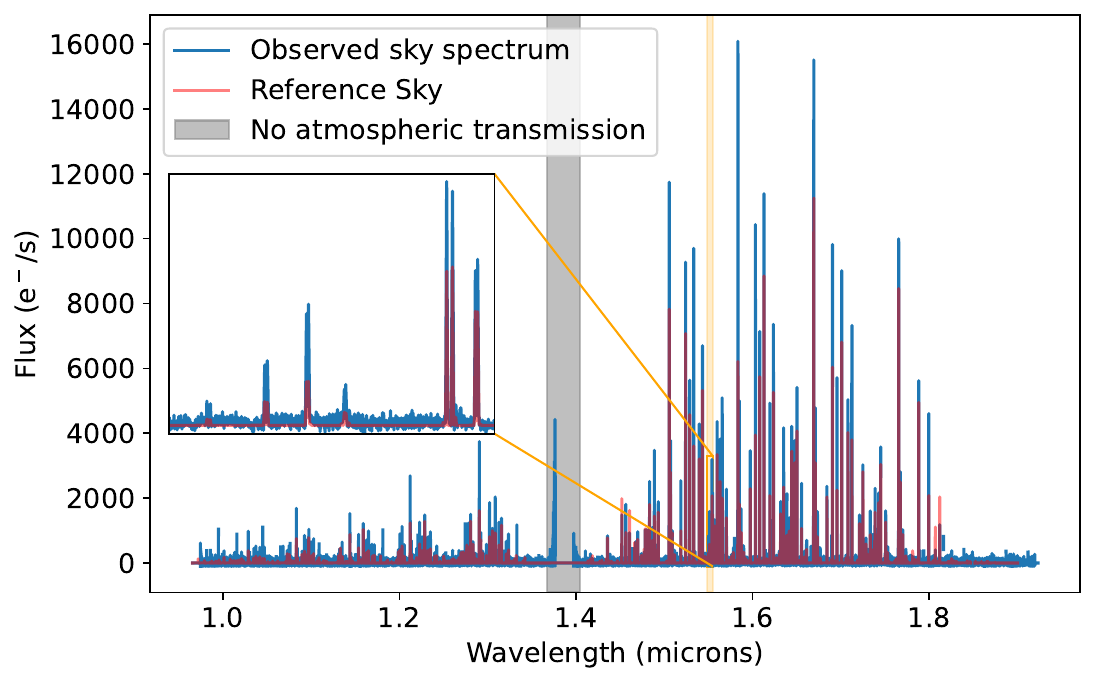}
\centering
\caption[]{Single NIRPS exposure of the night sky is presented in blue showcasing the emission line contamination. Flux is recorded in units of photo-electrons per second. The sky reference is plotted in red (Section \ref{ref_sky}) containing the emission lines corrected in the DRS workflow. The grey region corresponds to near-zero atmospheric transmission due to strong water absorption features. A zoom-in of the region in yellow  shows that a vast majority of emission features are detected by our algorithms.}
\label{s1d_emission}
\end{figure}

Another characteristic of the atmospheric emission lines is that their intensity varies depending on the atmospheric conditions, airmass, and time of observation~\citep{dauphin2024}. This effect is particularly pronounced during twilight, when high-energy solar photons excite additional vibrational modes. Consequently, relying on information from a single sky exposure to correct all frames would lead to inconsistencies. Since the emission lines are a consequence of the Earth's atmosphere, their location will remain fixed in the detector (Earth) frame of reference, but if the wavelength grid is shifted to the barycentric frame (Solar System frame), they will shift on the wavelength grid depending on the barycentric Earth RV (BERV) across all epochs, while the stellar lines will shift in accordance with the systematic velocity of the star with respect to the Solar System. All of these factors indicate that simply masking the emission lines or modifying the CCF mask would not be sufficient or feasible to mitigate their impact. For EPRV, any such excursions in the spectrum will degrade the retrieved RV precision; therefore, an efficient correction needs to be implemented to realise the full capabilities of NIRPS (or any NIR instrument).

\subsection{Correction methods in the literature}

There have been several IR ground-based spectrographs introduced prior to NIRPS and, naturally, they have also been aimed at addressing the same issues related to the Earth's atmosphere. Different observing strategies and techniques have been developed and used to varying levels of success. In CRIRES\footnote{CRIRES+ user manual, Document no.: ESO-254264, https://www.eso.org/sci/facilities/paranal/instruments/crires/doc.html} \citep{crires_paper}, the technique of `nodding' \citep{nodding_ref} is used. It involves observing a target through two different points along a slit, in several subexposures. Then by subtracting one from the other, the background emission can be removed. GIANO \citep{giano_paper} has two fibres that can observe the target star and the sky background simultaneously. Since both fibres are of the same size (1\arcs\ diameter), the sky background is removed by subtracting the flux from the second fibre \citep{giano, giano_cookbook}. A theoretical approach to model the emission spectrum \citep{oh_modelling} is also explored. CARMENES \citep{carmenes_paper} also takes advantage of its double fibre configuration by simultaneously observing the target star and the sky background. Reference spectra for both a target and its sky background are first computed using all the exposures, then the sky reference is scaled to match the intensity of the lines seen in the stellar spectrum. This scaled background is then subtracted to remove the atmospheric emission features, as demonstrated in \cite{carmenes}. If limited by readout noise, this technique contributes a factor of $\sqrt{2}$  to the noise across the entire spectrum which, in turn, increases the telescope time by a factor of 2 to retain the same signal-to-noise ratio (S/N).

\section{NIRPS sky subtraction} \label{sec3}

\subsection{Fibre geometry and implication for sky subtraction}
As mentioned above, NIRPS has the advantage of using two optical fibres in its observations. Fibre A observes the target, generally a star, and fibre B acts as the calibration fibre and observes the sky. Observing in HA mode features the use of 0.4\arcsec fibres (octagonal image on detector) for both the science and calibration channels. So, one can perform a simple local subtraction of the simultaneous sky calibration from the observed stellar spectrum to remove the telluric emission features. On the other hand, the HE science fibre is larger (0.9\arcsec on-sky as a rectangular image on the detector) to maximise the instrument throughput, but the corresponding sky fibre (0.4\arcsec on sky as an octagonal image on the detector) does not have the same format. This is because it would be impossible to package the diffraction orders onto the detector without having overlapping traces. Therefore, fibres A and B have a varying spectral resolution, so we cannot simply subtract the flux of the sky calibration fibre from the science frames to obtain the sky subtracted frames.

Due to the difference in on-sky area between the two fibres (for HE mode), the sky calibration fibre receives about five times less flux than the science fibre. Even without consideration of resolution mismatch, scaling the sky fibre flux onto the science channel would increase readout noise five-fold in the sky subtracted frames. The approaches taken by the NIRPS DRS and APERO DRS yield different solutions to this common challenge; the sky spectrum cannot be naively subtracted using a simple scaling. The methods revolve around a core idea: one constructs a very high-S/N normalised spectrum (deep-stack) composed of only emission features detected in sky observations and, at the per-line level, these get subtracted from the science frame. This approach has the advantage of not contributing noise in domains that are devoid of sky lines and photon noise on the deep-stack is minimised. Since the correction algorithms were designed to operate independent of any resolution differences between the fibres, they are equally effective in HA mode, where the science and calibration fibres are identical.

\subsection{Reference sky spectrum} \label{ref_sky}

Technical nights for the maintenance of the instrument occur every few months and, during these nights, night-time sky-sky (sky on both fibres) exposures were obtained for calibration purposes. These correspond to 6$\times$10-minute exposures (60 minutes in total) each night. So far, we have obtained a total of 66 sky-sky frames for HE mode across 11 different nights and 24 for HA mode from 4 different nights, recorded over several months (January 2023-November 2024).

In principle, a sky-sky frame taken immediately before a science exposure could be subtracted directly, but the sky varies on short timescales (tens of minutes; \citealt{noll2023, dauphin2024}) and for faint targets requiring long exposures (e.g. TESS follow-ups), which are the most sensitive to sky contamination (Fig.~\ref{emission_jmag}), this method fails. Moreover, individual sky-sky pairs lack a sufficient S/N to reliably detect emission lines, which also vary with atmospheric conditions.

To address this, we created high-S/N, deep-stack reference sky spectra by combining sky-sky frames collected over several months. Four such spectra are used in the NIRPS DRS: HE mode reference A and B; and HA mode reference A and B. Since APERO does not use the calibration fibre, only HE/HA reference A spectra were created. The construction process is the same across pipelines, modes, and fibres: sky-sky frames were separated by spectral order, interpolated onto a common wavelength grid using the 3D spline function in \texttt{scipy} \citep{2020SciPy-NMeth}, and combined via a weighted median corresponding to per-frame S/N.

\subsection{line selection and identification}

Sky emission lines are located using a high-pass filter, implemented with the \texttt{pandas} rolling median and MAD per spectral order \citep{pandas1, pandas2}. MAD is preferred over standard deviation as it is less sensitive to outliers, in this case the emission features. A window of 100 pixels was used, with pixels $>6\times$MAD above the rolling median flagged as potential lines, balancing noise rejection and line detection.

This method can also flag spurious features from detector systematics, readout errors, or cosmic rays and, thus, further filtering needs to be applied. Emission lines appear as singlets, doublets, or blended doublets; Gaussian fits with \texttt{lmfit} \citep{lmfit} are used to estimate their full width at half maximum (FWHM). Candidates with a FWHM smaller than the spectral resolution or much larger than twice it are rejected. A region of $\pm$1.5 FWHM was then defined around each line where the correction would be localised. Blended or closely spaced doublets were merged into single regions to avoid overlaps in the correction algorithm.

In HE mode, fibre A admits more flux and thus more lines than fibre B, but only 293 regions are common to both references A and B. The final line list (Table~\ref{linelist_sneakpeek}) contains 555 lines, mostly OH and O$_2$: 445 unblended OH, 67 unblended O$_2$, 15 blended OH, 4 blended OH+O$_2$, and 22 from other species, identified with the \texttt{HITRAN} catalogue \citep{hitran}. A comparison with previously published line lists from \citet{oh_modelling} and \citet{dauphin2024} is also presented.

\begin{figure*}[h]
    \centering
    \begin{subfigure}[t]{0.49\linewidth}
        \centering
        \includegraphics[width=\linewidth]{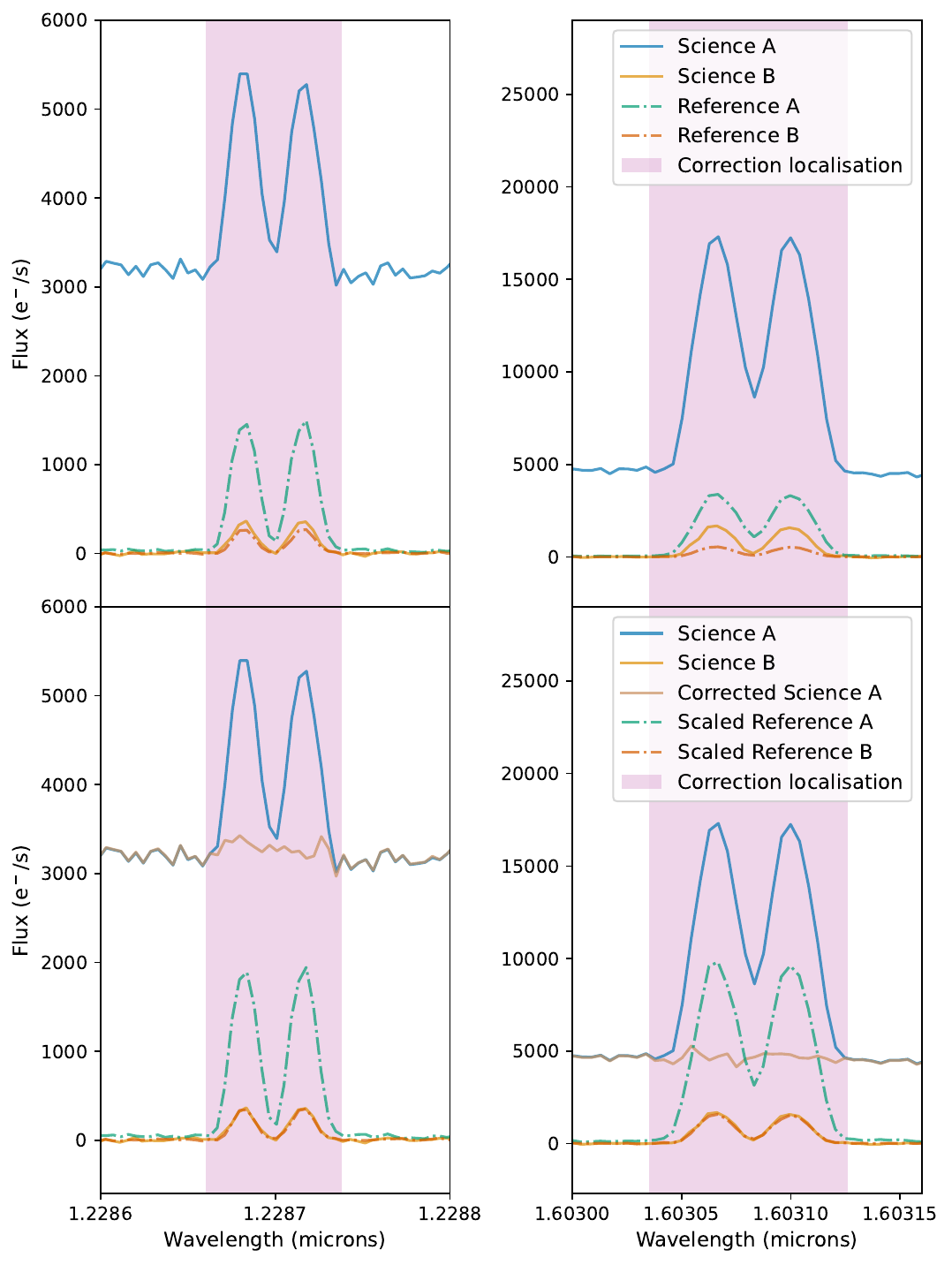}
        \caption{NIRPS DRS correction.}
    \end{subfigure}
    \hfill
    \begin{subfigure}[t]{0.49\linewidth}
        \centering
        \includegraphics[width=\linewidth]{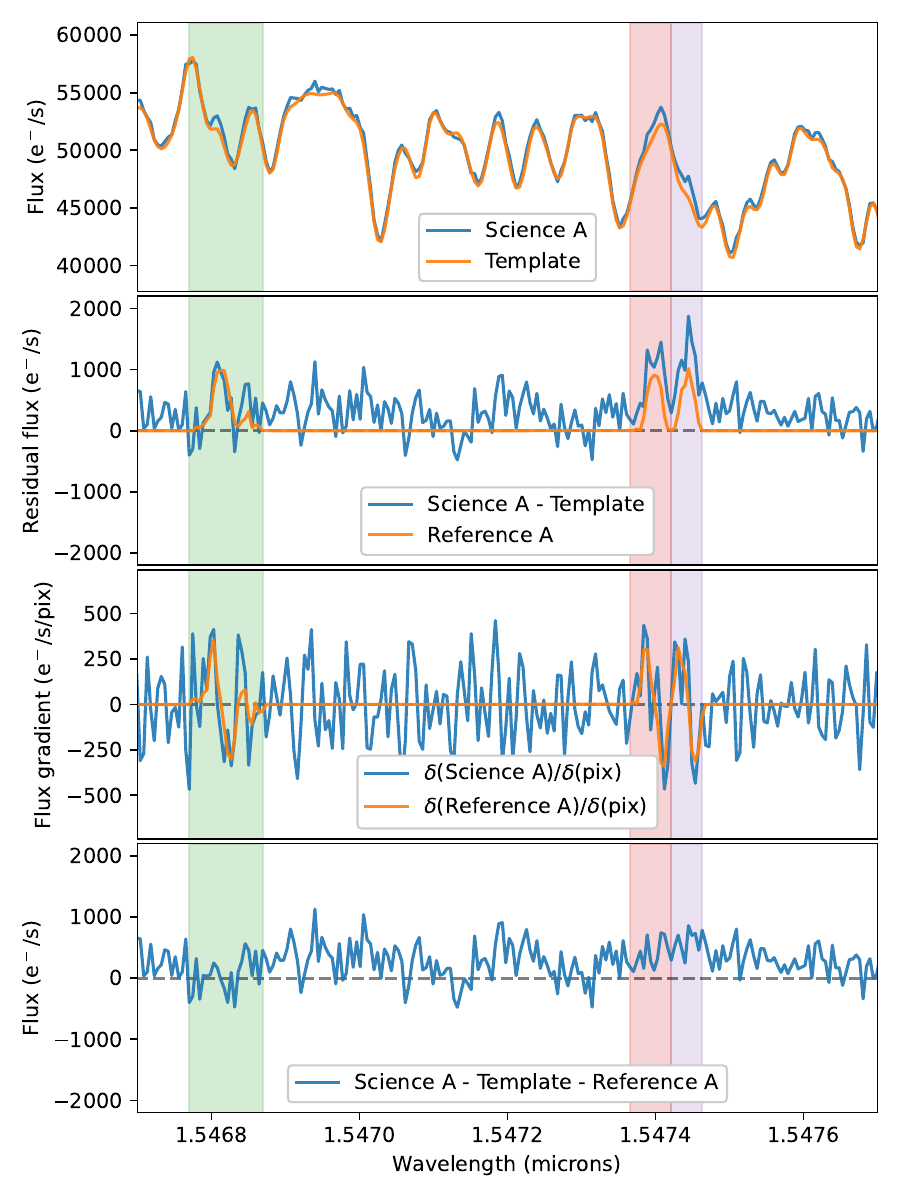}
        \caption{APERO DRS correction.}
    \end{subfigure}
    \caption{(a) Emission line correction performed on a NIRPS DRS spectrum of TOI-406 \citep{lacedelli2024} at two different doublet features, 1.2287 $\mu$m and 1.603055 $\mu$m, using the sky background method. The correction on the same frame and over the same wavelength range as done by APERO DRS is demonstrated in the Appendix (Fig. \ref{apero_oh_corr_demo_blue}) (b) APERO DRS correction technique on a singlet (1.5469 $\mu$m) and a doublet (1.54743 $\mu$m) emission line in the spectrum of a bright target, Proxima Centauri \citep{asm2025}. 'A' and 'B' refer to the science and calibration fibres, 'science' refers to the spectra obtained during a science observation, 'reference' refers to the high-S/N deep-stack spectrum of sky emission lines, 'pix' is short for pixels and 'template' refers to a median stellar template computed using all available observations of said target.}
    \label{fig:nirpsdrs_apero_demo}
\end{figure*}

\subsection{Sky subtraction in NIRPS DRS: Simultaneous sky background technique}

The implementation in the NIRPS DRS (since DRS-3.2.0), uses the simultaneous sky background calibration obtained during a science observation as a relative scaling measure for the emission lines to be subtracted. Assuming that atmospheric conditions do not change drastically over the 37\arcs\ separation between the two fibres on the sky, and that the relative throughput (flux) between fibre A and fibre B does not evolve significantly over time, this relative scaling factor can be applied to reference A before locally subtracting from the science A spectrum. The exact process is outlined below as well and detailed in Fig. \ref{fig:nirpsdrs_apero_demo}:
\begin{enumerate}

    \item For each frame, the 2D uncorrected target spectrum (science A) and the simultaneous background calibration spectrum (science B) are selected. Emission lines detected in both, science B and reference B are then identified as the common set. \\

    \item Only for these common lines, the per-line flux ratio between science B and reference B is calculated by summing the net flux in their respective line region (integrated flux). A  `median ratio' is then computed across all these lines. This median ratio is then used to correct for the fainter lines seen in science A but not in science B. \\

    \item Next, all the lines are sorted into two categories. Ones that would be corrected using their individual flux ratios (per-line ratios) and others that would be corrected using the median ratio for those only detected in science A. An additional check is done on the per-line ratio to make sure they remain physically plausible, if not, then that line is corrected using the median ratio.\\

    \item To perform the correction, the 1.5 FWHM region around a line is taken from the list and the closest starting and ending wavelengths are located on the science A wavelength solution. Then the reference A line region is interpolated to match the wavelength grid of previously obtained line region in science A. The flux on each pixel in the reference A line region is scaled by the median or per-line ratio (depending on the line) before subtracting this new flux from the science A spectrum following:
    
\begin{equation} \label{skycorrectioneqn}
\text{Science A Corrected} = \text{Science A} - \text{Ratio} \times \text{Reference A}
\end{equation}

    \item This process is repeated for every line in each spectral order as well as every new frame because the per-line ratios and number of common lines can differ significantly depending on the atmospheric conditions of the night. The only static data products are: reference A, reference B spectra and the line list. 
    
\end{enumerate}

\subsection{Sky subtraction in APERO DRS: Derivative technique}

The versatility of APERO DRS allows it to utilise all available observations of a particular target during the reduction process when available, facilitating the derivative technique for sky subtraction. The current APERO implementation (v0.7.293) for NIRPS uses a least-squares fit of the flux derivative (rather than the flux itself) to scale the emission lines. It offers a significant advantage as when subtracting the stellar template (a necessary step here) from the observation of interest, there are low-level differences that arise from slight differences in the injection and modal noise. A least-squares fit of the difference in flux would therefore be impacted by continuum mismatch, while the derivative of the flux is minimally affected. This technique does not rely on the reference B spectrum (only reference A) and it can therefore be applied to instruments such as SPIRou, which do not perform simultaneous sky calibration. 

The exact steps for the procedure are highlighted below and demonstrated in Fig. \ref{fig:nirpsdrs_apero_demo}, although only the scaling mechanism differs from the NIRPS DRS procedure:

\begin{enumerate}

    \item A high-S/N stellar template is first constructed by taking all telluric absorption-corrected observations of the target star and computing their median. If the spectra are obtained at different BERVs, any residual telluric emission features are effectively averaged out, preventing contamination of the template. A 2D science frame (science A) is then selected, and the median stellar template is subtracted, leaving only telluric features. For each emission line in reference A, the corresponding line is identified in the template-subtracted science A, and reference A is interpolated onto the wavelength grid of science A. \\

    \item For both spectra, the per-pixel derivative with respect to wavelength is computed as  
    
\begin{equation} \label{derivative}
    \frac{\mathrm{d}}{\mathrm{d}\lambda}F_n = \frac{F_n - F_{n-1}}{\lambda_n - \lambda_{n-1}},
\end{equation}  

    where $n$ is the pixel index. A least-squares fit then scales the reference A derivative to the science A derivative, yielding the per-line ratio used to adjust reference A before subtracting it from science A. \\

    \item This is repeated for all lines in the reference A library for each spectral order, so that every emission feature is corrected by its own per-line ratio. Finally, the stellar spectrum is re-added. \\

\end{enumerate}

\begin{figure}[h]
\includegraphics[width=\columnwidth]{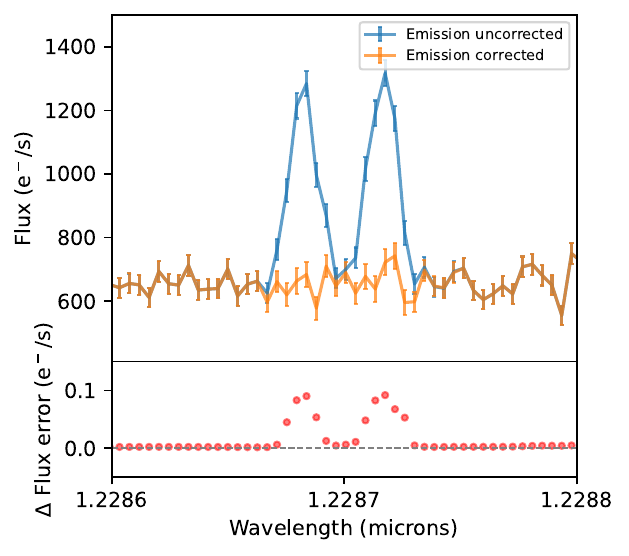}
\centering
\caption[]{Demonstration of the emission line correction on a faint target TOI-3494 observed in HA mode. Here, the bottom figure presents the difference in the per-pixel error bars before and after the correction. Following the error propagation highlighted in Section \ref{error_propagation}, the increase in the per-pixel error is localised to the location of the emission features.}
\label{toi3494_error}
\end{figure}

\subsection{Error propagation} \label{error_propagation}

Since the flux in the science spectrum is altered locally during the process of the emission line correction, it is important to correctly inflate the error bars as well to reflect this change. Line regions with strong emission lines should have relatively larger error bars because the errors from the reference sky spectrum need to be propagated into the emission corrected spectra. The following propagation method can be used:

\begin{enumerate}

    \item Each frame that goes into the creation of the reference sky spectrum has the same exposure time, which implies that their S/N should be comparable within the limits of the sky variation. This is expressed as

    \begin{equation} \label{mastererr}
    \sigma_{reference} = \left( \sum \frac{1}{\sigma_{frame}^2} \right)^{-1/2}.
\end{equation}
    
This equation gives the calculation of the error for each pixel of the reference sky spectrum ($\sigma_{reference}$) with $\sigma_{frame}$ being the error bar of that pixel for a given frame. This is in theory not an exact propagation because of the individual frames being interpolated to match a common wavelength grid before the reference sky spectrum was created, but since the pixel scale is roughly 1\,km/s, which is small enough compared to the width of a typical emission line ($\sim$4\,km/s), no significant bias is introduced. The comparison relative to the continuum noise is presented in Fig. \ref{toi3494_error}.

    \item As the correction process is a local subtraction, the net per-pixel error of the final corrected spectrum would amount to the error of the two spectra added in quadrature keeping in mind the scaling factor, $R$, used for reference A. The calculation of the errors of the corrected spectrum, $\sigma_{corr}$, follows

    \begin{equation} \label{correrr}
    \sigma_{corr}\, =\, \sqrt{\sigma_{science}^2\, +\, (\sigma_{reference}*R)^2},
\end{equation}
    
    where $\sigma_{science}$ is the error for each pixel of science A.

\end{enumerate}

\subsection{Justification for empirical-based correction}

While previous works \citep{oh_modelling} have explored the possibility to model the night sky airglow lines to better correct them, we developed empirical-based correction techniques for NIRPS to enable fast, straightforward data reduction at the telescope. 
A fully model-based treatment, similar to telluric absorption correction, could provide incremental performance gains but typically requires substantially longer processing times.
Since telluric emission features are fewer in number compared to absorption lines and have the biggest impact on faint targets (Fig. \ref{emission_jmag}), we only corrected the few hundreds of lines that impact RV measurements. The LBL method of RV extraction is outlier-resilient; hence, the fainter emission lines buried in the photon noise are automatically rejected during the RV calculation. Additionally, the lack of an atmospheric model allows for the algorithms to be more flexible in implementation to other possible instruments such as ANDES at ELT~\citep{marconi2022}. This is because the line list is constructed using the instrument's sky observations and, hence, only the resolved lines are corrected. A validation of our techniques (Section \ref{sec4}) showcases that our empirical-based approach performs adequately.

\section{Validation on NIRPS targets} \label{sec4}

\subsection{Visual inspection of the spectrum}

To demonstrate that the correction works effectively for the full range of observable targets, two extreme cases need to be tested. As illustrated in Fig. \ref{fig:nirpsdrs_apero_demo}, the two algorithms of the emission correction are given for faint target (NIRPS DRS: TOI-406, \citealp{lacedelli2024}) and a bright target (APERO DRS: Proxima, \citealt{asm2025}). In the case of Proxima, the emission doublet is blended into the stellar spectrum. The opposite is true for TOI-406 where the emission features can be clearly distinguished from the stellar continuum visually. In both cases, as explained in Section \ref{sec3}, the correction is localised to the pixels affected by the emission lines, leaving the continuum unaltered.

An alternative method of measuring the effectiveness of the correction is to utilise the CCF technique, where instead of using a mask composed of stellar lines, we can create a mask consisting of emission lines from the line list. In Fig.~\ref{toi406-ccf} this CCF mask is used on a frame of TOI-406 resulting in the peak in the Earth's rest frame at that epoch. The contrast shown here depicts the average intensity of the emission lines for this spectrum relative to the continuum. After the emission correction, the resulting CCF is nearly flat, with residuals of the order of 1\% or less, as indicated by the contrast of the CCF peak. The same test is repeated on a frame of TOI-4552~\citep{Srivastava2026} and presented in the appendix (Fig.~\ref{toi4552-ccf}).

\begin{figure*}[h]
    \centering
    \begin{subfigure}{0.48\textwidth}
         \centering

         \includegraphics[width=\linewidth]{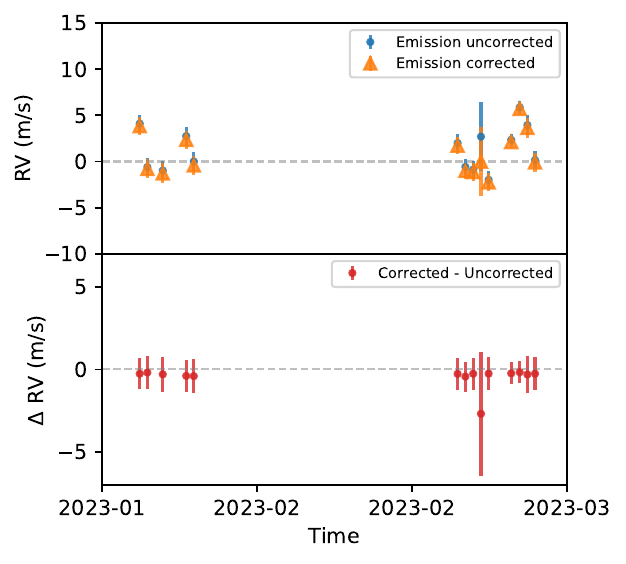}
                  \caption{Cross-correlation function RVs}
    \end{subfigure}
    \hfill
    \begin{subfigure}{0.48\textwidth}
         \centering

         \includegraphics[width=\linewidth]{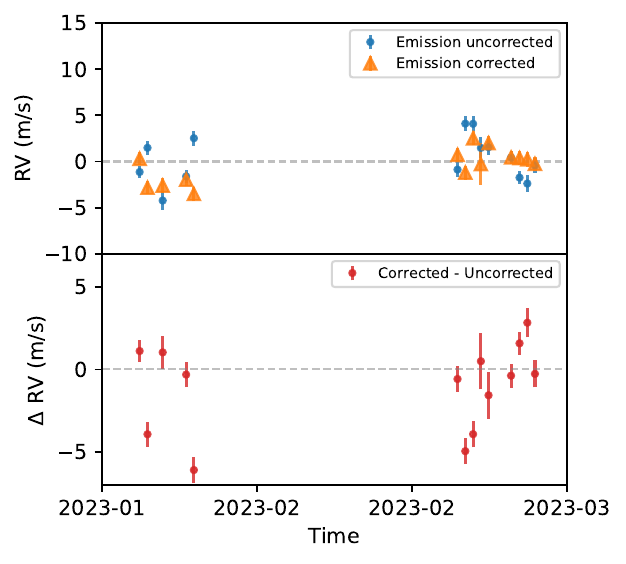}
                  \caption{Line-by-line RVs}
    \end{subfigure}
    \caption{(a) Comparison of the CCF RVs derived using NIRPS DRS pipeline operating at the telescope site for emission corrected and uncorrected spectra of Proxima Centauri. This data was recorded during the commissioning period of NIRPS. (b) Plot showcases the RV calculated using LBL for same spectra but processed with the APERO DRS. The `emission corrected' dataset is the one presented in \citet{asm2025}. The error bars in the residual plots are the maximum RV errors between the two datasets at that epoch as the datasets are correlated, the errors cannot be added in quadrature. Since LBL produces more precise RVs ($\sigma_{RV-CCF}=1.22\,$m/s, $\sigma_{RV-LBL}=0.80\,$m/s), the emission features need to be well corrected even for bright targets.}
    \label{proxima_RV}
\end{figure*}

\subsection{Effect on extracted radial velocity}

Proxima Centauri is one of the `golden targets' of NIRPS and it has been extensively observed in both HA and HE mode, making it a prime candidate to validate the correction algorithm. We compared the extracted RV measurements before and after emission correction on the Proxima spectra obtained during the instrument commissioning period (January 2023-March 2023) using both CCF and LBL algorithms. The CCF RVs presented here were extracted using the NIRPS DRS, while the LBL RVs come from APERO DRS, which were used in the final analysis in~\citet{asm2025}. The effect on both datasets can be seen pre and post emission line correction (see Fig.~\ref{proxima_RV}). LBL utilises more than ten times the number of stellar lines compared to the CCF approach to attain a much higher precision ($\sigma_{RV-LBL}$\,=\,0.80~m/s; $\sigma_{RV-CCF}$\,=\,1.22~m/s) and as such even minute uncorrected telluric residual features distort the line profiles of some stellar lines to an extent where the retrieved RVs are offset, contrary to the CCF extracted RVs.

\subsection{NIRPS DRS and APERO DRS correction comparison} \label{nirps_apero_compare}

\begin{figure}
\includegraphics[width=\columnwidth]{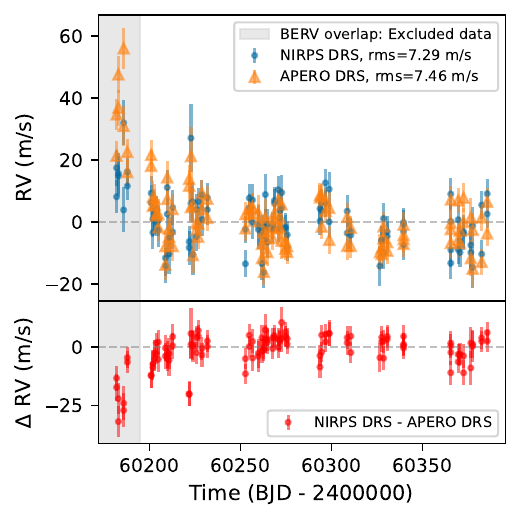}
\centering
\caption[]{TOI-406 RVs derived using LBL run on both NIRPS DRS and APERO DRS showcasing minimal difference in the extracted RVs. The error bars in the residual plot are the maximum values of the RV error between the two pipelines at that epoch. The planet TOI-406 b is recovered in both cases to within 1-$\sigma$ significance in agreement (see appendix Fig.~\ref{toi406_RV}). The grey area contains the datapoints excluded because of the BERV overlap effect discussed in Section~\ref{masking_code}. The APERO DRS dataset is more significantly impacted compared to NIRPS DRS dataset.}
\label{pipe_compare}
\end{figure}

An ideal faint target to test the validity of the algorithm representative of the sample of targets in the TESS follow-up NIRPS subprogramme is TOI-406, already published in \citet{lacedelli2024} with a significant detection combining NIRPS, HARPS, and ESPRESSO. The published RVs were extracted using the APERO DRS framework; however, we demonstrate here that both NIRPS DRS and APERO DRS produce results that agree within a 1-$\sigma$ significance (Fig. \ref{pipe_compare}). The shaded area includes the observations carried out at an epoch where the systemic velocity ($v_{sys}$) of the target and the BERV overlapped within a window of 8~km/s, resulting in imperfect telluric absorption and emission corrections causing a scatter in the extracted RVs. This effect and ways to mitigate it are discusses in detail in Section~\ref{masking_code}. The recovered semi-amplitudes for TOI-406 b are: $3.76 \pm 0.81$ m/s for APERO DRS and $3.77 \pm 0.91$ m/s for NIRPS DRS (Appendix: Fig. \ref{toi406_RV}). Given the differences in the methodology involved in the correction of telluric emission features, we should expect some subtle differences. Nonetheless, the overall effectiveness of both techniques is expected to be comparable.

\section{Discussion} \label{sec5}

\subsection{Sequence of telluric absorption and emission corrections}

Removing Earth's atmospheric contamination from a NIR spectrum involves two separate corrections, emission and absorption, which are performed using different methods. Ideally, both should be applied together to avoid compounding errors; however, in practice they must be done sequentially, making the order important.

In NIRPS DRS, emission features are corrected first because: the reference sky spectrum was built from uncorrected telluric absorption observations; this order maintains consistency with the ESPRESSO DRS workflow, where the sky subtraction precedes telluric absorption correction \citep{espresso,allart2022}; and telluric emission is additive, while absorption is multiplicative, so emission must be subtracted before absorption alters the continuum. In contrast, APERO DRS corrects emission after correcting absorption, since constructing the stellar spectrum for emission removal requires an initial telluric absorption correction. 

\subsection{Effect of telluric residuals on EPRV} \label{masking_code}

As showcased in Fig.~\ref{pipe_compare}, despite the correction algorithms implemented in both DRS, some telluric residuals remain due to incomplete line lists or over- or under-correction, resulting in RV offsets. These offsets imprint harmonics of the Earth’s orbital period (365/$n$~days) on the RV time series as they are an effect of the Earth's atmosphere. The effect is most pronounced when the BERV at observation is within two FWHM (i.e. FWHM$_{NIRPS-HE}\sim$4 km/s; FWHM$_{NIRPS-HA}\sim$3 km/s) of the star’s systemic velocity, $v_{sys}$, allowing for molecular lines that are common between the Earth's and M dwarfs' atmospheres (e.g. H$_2$O and OH;~\citealt{Rajpurohit2018}) to overlap. For very faint targets, these residuals remain significant relative to the stellar continuum (in flux), thereby distorting the line profiles and affecting the RVs. This BERV-overlap and the impact on RV timeseries has been discussed in a number of recently published works: \citet{Parc2025, Frensch2026, Srivastava2026, Osborn2026, Weisserman2026}.

In particular, the work by~\citet{Parc2025} offers a discussion of this effect and proposes one method of correcting the affected RVs, namely, by removing specific spectral lines from the LBL RV calculation that overlap with telluric emission features. However, such an approach cannot be universally applied, as the `problematic' lines may vary between targets and may not be affected to the same extent. To address this limitation, we developed two general methods to mitigate the residuals. The first involves masking regions of the spectrum that tend to exhibit noise excursions above a defined uncertainty threshold, cleaning the spectrum at the cost of RV precision. The second uses principal component analysis (PCA) to correct those regions, instead of excluding them.

The post-processing technique identifies and masks outliers in spectral time-series residuals. Each spectrum is first corrected for the BERV to align all spectra in a common rest frame. After normalisation, a median template is built by binning over BERV values to reduce uneven sampling effects. For each spectrum, the template is Doppler-shifted back to the observer frame, residuals (observed-minus-template) are computed, and high-frequency noise is suppressed with a low-pass filter to enhance outlier detection. Outliers are flagged at a chosen $\sigma$ level, with 1-$\sigma$ defined as half the distance between the 16th and 84th percentiles. Optionally, weighted PCA \citep{wpca} is applied to residuals in the OH reference frame to capture and subtract systematic trends. The output includes masked FITS files (outliers set to NaN) and summary products such as the sigma map, PCA components, and median template. As demonstrated in Fig.~\ref{berv_crossing}, this residual correction reduces the 75- and 90-day peaks (365/$n$ days) in the TOI-4552~\citep{Srivastava2026} time series below the false alarm probability threshold, confirming the successful telluric residual removal.

\section{Conclusion} \label{sec6}

One challenging aspect of high-resolution spectroscopy in the NIR is that Earth's atmosphere is less transparent, while also being brighter compared to the visible domain, making it necessary to develop tools that disentangle the spectral features of the Earth's atmosphere from the science observation. These telluric features are manifested in the form of absorption (CH$_4$, H$_2$O, CO$_2$, O$_2$) and emission (OH, O$_2$) lines that span the entire 0.98 $\mu$m to 1.9 $\mu$m wavelength coverage of NIRPS and beyond. Here, we present two empirical-based algorithms used to correct for the telluric emission features, as employed in the data reduction frameworks of NIRPS DRS and APERO DRS. The principles behind the algorithms are instrument-independent and can be applied to any high-resolution spectrograph, such as ANDES at ELT.

We have validated our results on both bright (Proxima) and faint (TOI-406, TOI-3494) targets and and demonstrated that despite the different algorithms employed in the two DRS, the results display a strong agreement with respect to both the extracted RV performance and visual inspection of the spectrum. Currently, we can correct for 445 non-blended OH lines, 67 non-blended O$_2$ molecular lines, 15 blended OH lines, 4 blended OH and O$_2$ lines, and 22 lines from species that are neither OH or O$_2$ (Table \ref{linelist_sneakpeek}). These corrections leave residuals at the $\sim$1\% level, allowing for the performance of NIRPS in the $\le$1~m/s domain.

\section{Data availability}

The entirety of Table~\ref{linelist_sneakpeek} and reference sky spectra for both HA and HE instruments modes and both fibres A and B reduced by both APERO DRS and NIRPS DRS are available on Zenodo via the following DOI: \href{https://doi.org/10.5281/zenodo.21894247}{https://doi.org/10.5281/zenodo.21894247}.

\begin{acknowledgements}
XDu  acknowledges the support from the European Research Council (ERC) under the European Union’s Horizon 2020 research and innovation programme (grant agreement SCORE No 851555) and from the Swiss National Science Foundation under the grant SPECTRE (No 200021\_215200).\\
This work has been carried out within the framework of the NCCR PlanetS supported by the Swiss National Science Foundation under grants 51NF40\_182901 and 51NF40\_205606.\\
\'EA, RD, NJC, FBa, LMa, BB, RA, CC, PL, TV \& JPW  acknowledge the financial support of the FRQ-NT through the Centre de recherche en astrophysique du Qu\'ebec as well as the support from the Trottier Family Foundation and the Trottier Institute for Research on Exoplanets.\\
\'EA, RD, FBa \& LMa  acknowledges support from Canada Foundation for Innovation (CFI) program, the Universit\'e de Montr\'eal and Universit\'e Laval, the Canada Economic Development (CED) program and the Ministere of Economy, Innovation and Energy (MEIE).\\
SCB, ED-M \& NCS  acknowledge the support from FCT - Funda\c{c}\~ao para a Ci\^encia e a Tecnologia through national funds by these grants: UIDB/04434/2020, UIDP/04434/2020.\\
SCB   acknowledges the support from Funda\c{c}\~ao para a Ci\^encia e Tecnologia (FCT) in the form of a work contract through the Scientific Employment Incentive program with reference 2023.06687.CEECIND and DOI \href{https://doi.org/10.54499/2023.06687.CEECIND/CP2839/CT0002}{10.54499/2023.06687.CEECIND/CP2839/CT0002.}\\
XB \& XDe  acknowledge funding from the French ANR under contract number ANR\-24\-CE49\-3397 (ORVET), and the French National Research Agency in the framework of the Investissements d'Avenir program (ANR-15-IDEX-02), through the funding of the ``Origin of Life" project of the Grenoble-Alpes University.\\
NBC  acknowledges support from an NSERC Discovery Grant, a Canada Research Chair, and an Arthur B. McDonald Fellowship, and thanks the Trottier Space Institute for its financial support and dynamic intellectual environment.\\
ED-M  further acknowledges the support from FCT through Stimulus FCT contract 2021.01294.CEECIND. ED-M  acknowledges the support by the Ram\'on y Cajal contract RyC2022-035854-I funded by MICIU/AEI/10.13039/501100011033 and by ESF+.\\
DE  acknowledge support from the Swiss National Science Foundation for project 200021\_200726. The authors acknowledge the financial support of the SNSF.\\
JIGH, RR, ASM, NN \& AKS  acknowledge financial support from the Spanish Ministry of Science, Innovation and Universities (MICIU) projects PID2020-117493GB-I00 and PID2023-149982NB-I00.\\
The Board of Observational and Instrumental Astronomy (NAOS) at the Federal University of Rio Grande do Norte's research activities are supported by continuous grants from the Brazilian funding agency CNPq. This study was partially funded by the Coordena\c{c}\~ao de Aperfei\c{c}oamento de Pessoal de N\'ivel Superior—Brasil (CAPES) — Finance Code 001 and the CAPES-Print program.\\
ICL  acknowledges CNPq research fellowships (Grant No. 313103/2022-4).\\
BLCM  acknowledge CAPES postdoctoral fellowships.\\
BLCM  acknowledges CNPq research fellowships (Grant No. 305804/2022-7).\\
CM  acknowledges the funding from the Swiss National Science Foundation under grant 200021\_204847 “PlanetsInTime”.\\
Co-funded by the European Union (ERC, FIERCE, 101052347). Views and opinions expressed are however those of the author(s) only and do not necessarily reflect those of the European Union or the European Research Council. Neither the European Union nor the granting authority can be held responsible for them.\\
JRM  acknowledges CNPq research fellowships (Grant No. 308928/2019-9).\\
GAW is supported by a Discovery Grant from the Natural Sciences and Engineering Research Council (NSERC) of Canada.\\
RA  acknowledges the Swiss National Science Foundation (SNSF) support under the Post-Doc Mobility grant P500PT\_222212 and the support of the Institut Trottier de Recherche sur les Exoplan\`etes (IREx).\\
This project has received funding from the European Research Council (ERC) under the European Union's Horizon 2020 research and innovation programme (project {\sc Spice Dune}, grant agreement No 947634). This material reflects only the authors' views and the Commission is not liable for any use that may be made of the information contained therein.\\
0\\
KAM  acknowledges support from the Swiss National Science Foundation (SNSF) under the Postdoc Mobility grant P500PT\_230225.\\
NN  acknowledges financial support by Light Bridges S.L, Las Palmas de Gran Canaria.\\
NN acknowledges funding from Light Bridges for the Doctoral Thesis "Habitable Earth-like planets with ESPRESSO and NIRPS", in cooperation with the Instituto de Astrof\'isica de Canarias, and the use of Indefeasible Computer Rights (ICR) being commissioned at the ASTRO POC project in the Island of Tenerife, Canary Islands (Spain). The ICR-ASTRONOMY used for his research was provided by Light Bridges in cooperation with Hewlett Packard Enterprise (HPE).\\
AKS  acknowledges financial support from La Caixa Foundation (ID 100010434) under the grant LCF/BQ/DI23/11990071.\\
TV  acknowledges support from the Fonds de recherche du Qu\'ebec (FRQ) - Secteur Nature et technologies under file no. 320056.
\end{acknowledgements}

\bibliographystyle{aa}
\bibliography{biblitex}

\begin{appendix}
\onecolumn

\section{Supplementary plots}

\begin{figure}[H] 
\includegraphics[width=.62\columnwidth]{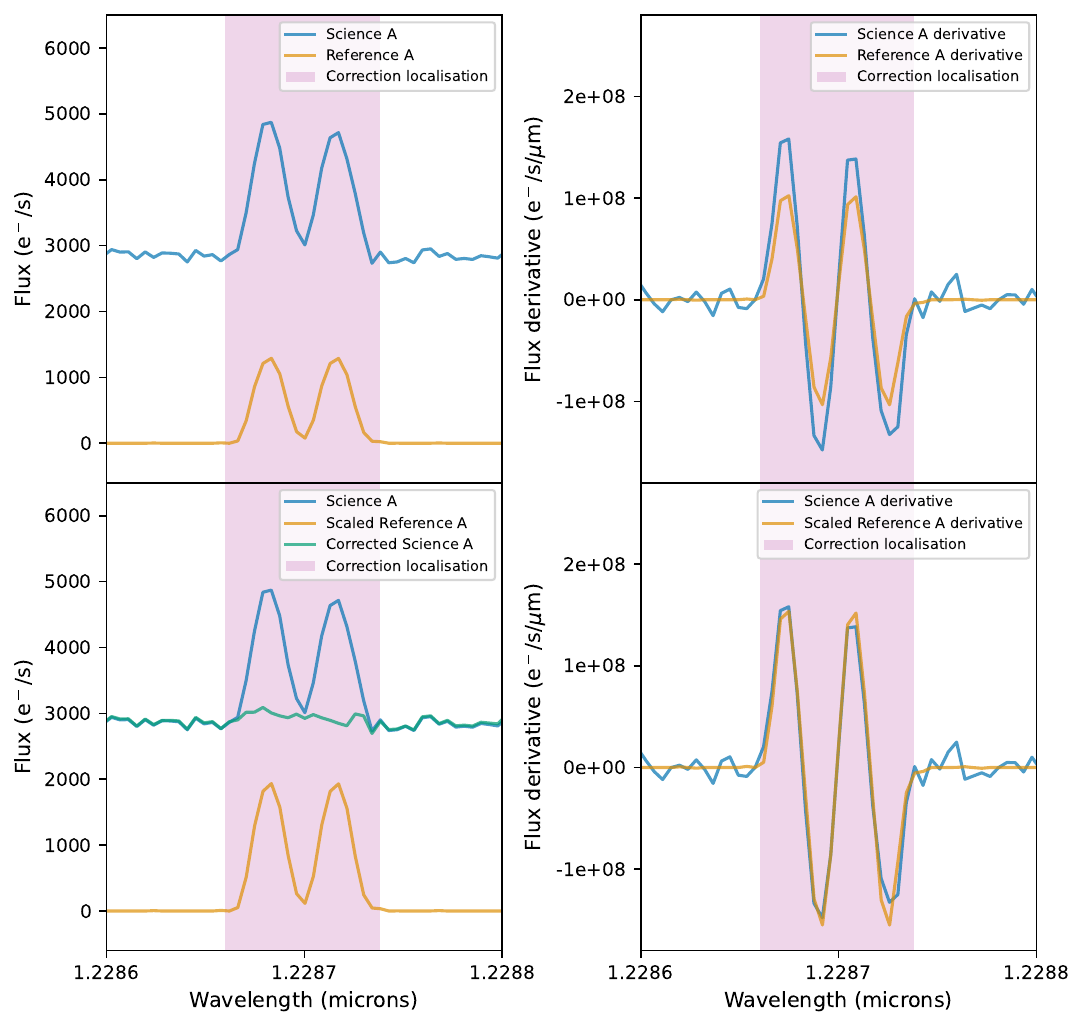}
\centering
\caption[]{Demonstration of the APERO DRS emission line correction for the same 1.2287 $\mu$m doublet in the same spectrum of TOI-406 as showcased in Fig. \ref{fig:nirpsdrs_apero_demo}. Both pipelines correct for the same feature at very comparable precision despite the different techniques employed.}
\label{apero_oh_corr_demo_blue}
\end{figure}

\begin{figure*}[h] 
    \centering
    \begin{subfigure}{0.49\textwidth}
         \centering
         \includegraphics[width=\linewidth, trim=0 0 0 0, clip]{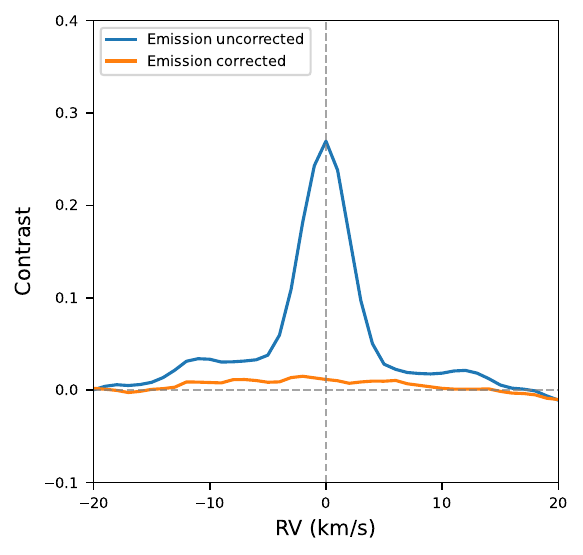}
         \subcaption{TOI-406}
         \label{toi406-ccf}
    \end{subfigure}
    \begin{subfigure}{0.49\textwidth}
         \centering
         \includegraphics[width=\linewidth, trim=0 0 0 0, clip]{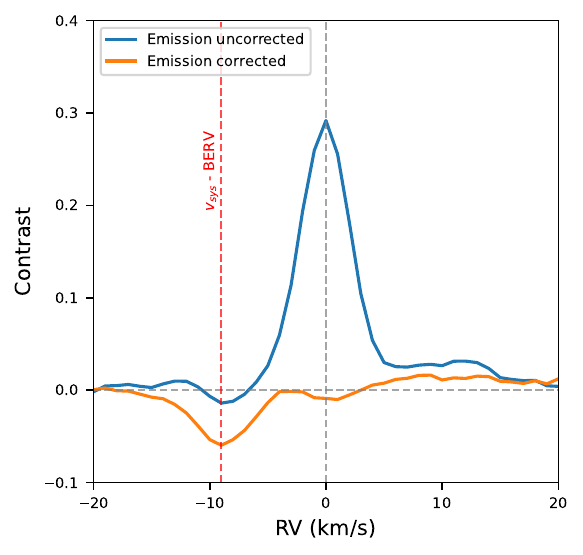}
         \subcaption{TOI-4552}
         \label{toi4552-ccf}
    \end{subfigure}
    \caption{CCF mask constructed using the detected emission lines is used on a frame of TOI-406 (left) and TOI-4552 (right) before and after it is processed by the NIRPS DRS emission correction. The CCF peaks at 0~km/s as the spectrum has been shifted to the Earth's rest frame. Post-correction, the CCF contrast is significantly lower and leaves $\sim$1\% level of residuals. For TOI-4552, the $v_{sys}$ relative to the BERV is marked (in red) to showcase the imperfections in the correction algorithm that lead to residual telluric features.}
\end{figure*}

\begin{figure*}
    \centering
    \begin{subfigure}{0.49\textwidth}
         \centering
         \includegraphics[width=\linewidth]{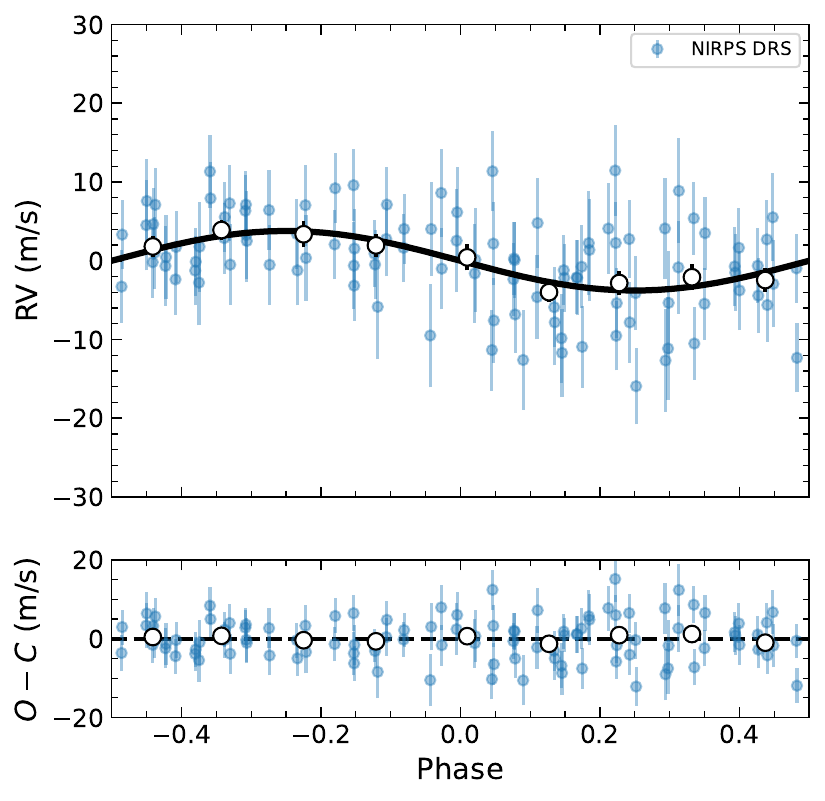}
    \end{subfigure}
    \hfill
    \begin{subfigure}{0.49\textwidth}
         \centering
         \includegraphics[width=\linewidth]{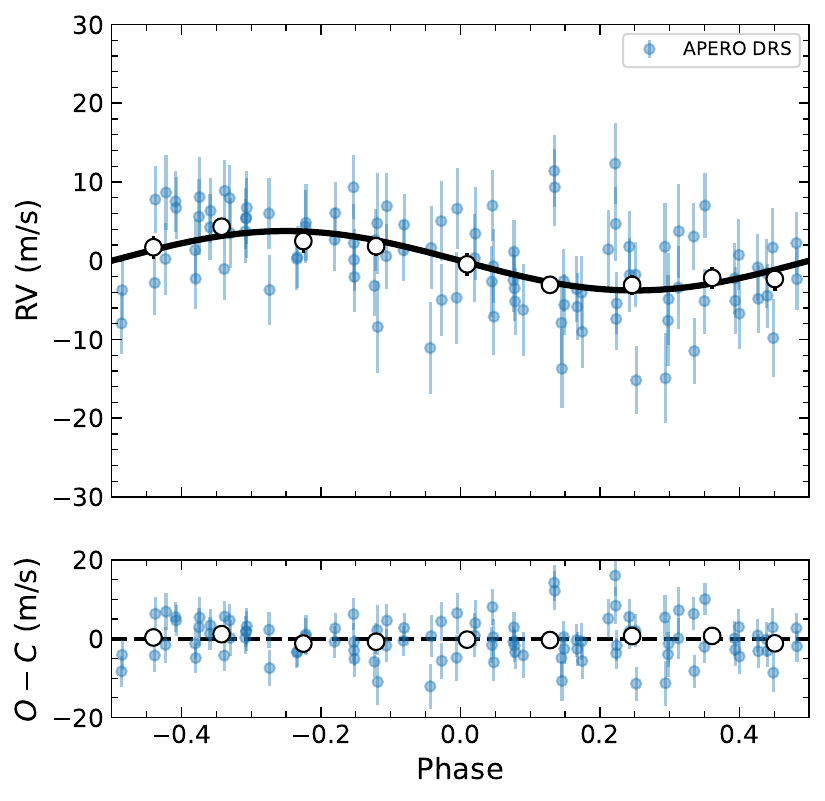}
    \end{subfigure}
    \caption{Comparison of LBL extracted RVs using the two DRS phase-folded to match the periodicity of TOI-406 b. To keep the comparison unbiased and similar to the published result, the telluric residual correction as discussed in Section \ref{masking_code} is not applied to either dataset. As depicted in Fig. \ref{pipe_compare}, a portion of the dataset is hindered by the BERV overlap and are thus excluded from this phasefold. The resulting semi-amplitudes are within 1-$\sigma$ agreement. $K_{NIRPS-DRS} = 3.77 \pm 0.91$ m/s, $K_{APERO-DRS} = 3.76 \pm 0.81$ m/s.}
    \label{toi406_RV}
\end{figure*}

\begin{figure*}
\centering
\includegraphics[width=\linewidth]{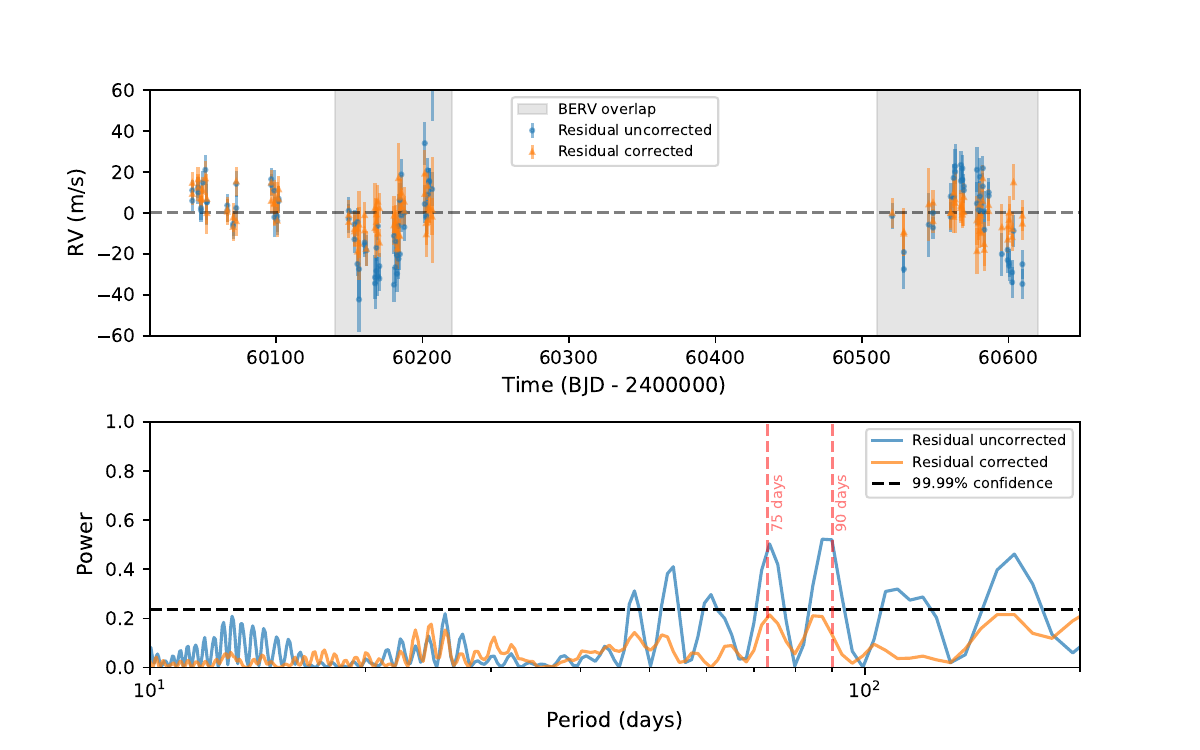}
\caption[]{Demonstration of the post-processing telluric residuals masking technique (Section~\ref{masking_code}) removing the harmonics of Earth's orbital period (90d and 75d; 356/$n$) that arise due to the BERV crossing effect. This timeseries is from a recently published NIRPS-GTO target TOI-4552~\citep{Srivastava2026}. Both the 90d and 75d peaks drop below the 0.01\% false alarm probability limit once we implement the residual correction algorithm, successfully removing the telluric signal.}
\label{berv_crossing}
\end{figure*}

\clearpage
\section{Tables}

\begin{center}
\begin{longtable}{@{}lllll@{}}
\caption{A shortened line list containing all the emission features detected and corrected in NIRPS spectra.}\\
\label{linelist_sneakpeek}\\
\hline
\toprule
\textbf{Wavelength ($\mu$m)} & \textbf{Chemical species} & \textbf{Blended} & \textbf{In \citet{oh_modelling}} & \textbf{In \citet{dauphin2024}}\\
\toprule
\endfirsthead
\toprule
\textbf{Wavelength ($\mu$m)} & \textbf{Chemical species} & \textbf{Blended} & \textbf{Oliva} & \textbf{Dauphin}\\
\toprule
\endhead
\bottomrule
\multicolumn{2}{r}{\small\itshape continued on next page} \\
\endfoot
\bottomrule
\endlastfoot
0.980255 & OH line & No & Yes & No \\
0.99115 & Unknown Species & No & No & No \\
0.991728 & OH line & No & Yes & No \\
0.994436 & OH line & No & Yes & No \\
0.994924 & OH line & No & Yes & No \\
0.995938 & OH line & No & Yes & No \\
1.001556 & OH line & No & Yes & Yes \\
1.003792 & Unknown Species & No & No & No \\
1.006336 & OH line & No & Yes & Yes \\
1.008523 & OH line & No & Yes & Yes \\
1.012692 & OH line & No & Yes & Yes \\
1.017431 & OH line & No & Yes & Yes \\
1.017501 & OH line & No & Yes & Yes \\
1.020583 & OH line & No & Yes & Yes \\
1.021161 & OH line & No & Yes & Yes \\
1.021383 & OH line & No & Yes & Yes \\
1.022834 & OH line & No & Yes & Yes \\
1.022887 & OH line & No & Yes & Yes \\
1.02874 & OH line & No & Yes & Yes \\
1.028944 & OH line & No & Yes & Yes \\
1.02987 & OH line & No & Yes & Yes \\
1.0299 & OH line & No & Yes & Yes \\
1.039937 & OH line & No & Yes & Yes \\
1.042106 & OH line & No & Yes & Yes \\
1.042139 & OH line & No & Yes & Yes \\
1.045337 & OH line & No & Yes & Yes \\
1.047161 & OH line & No & Yes & Yes \\
1.047199 & OH line & No & Yes & Yes \\
1.051207 & OH line & No & Yes & Yes \\
1.052735 & OH line & No & Yes & Yes \\
1.05752 & OH line & No & Yes & Yes \\
1.058834 & OH line & No & Yes & Yes \\
1.058913 & OH line & No & Yes & Yes \\
1.062362 & Unknown Species & No & No & No \\
1.07234 & OH line & No & Yes & Yes \\
1.073177 & OH line & No & Yes & Yes \\
1.074617 & OH line & No & Yes & Yes \\
1.075394 & OH line & No & Yes & Yes \\
1.077506 & OH line & No & Yes & Yes \\
1.08321 & OH line & No & Yes & Yes \\
1.083243 & OH line & No & Yes & Yes \\
1.083423 & OH line & No & Yes & Yes \\
1.084451 & OH line & No & Yes & Yes \\
1.084489 & OH line & No & Yes & Yes \\
1.084755 & OH line & No & Yes & Yes \\

\end{longtable}

\begin{tablenotes}
\item
\textbf{Notes:} The table contains the wavelength of the feature in the Earth rest frame, the chemical species responsible for the emission line and whether the lines were blended together or not. The line list is also compared to the lines identified in previous works: \citet{oh_modelling} and \citet{dauphin2024}. The full table is available on Zenodo.
\end{tablenotes}

\end{center}

\end{appendix}

\end{document}